\documentclass{article}
\usepackage{ijcai19}

\usepackage{times}
\usepackage{soul}
\usepackage{url}
\usepackage[hidelinks]{hyperref}
\usepackage[utf8]{inputenc}
\usepackage[small]{caption}
\usepackage{graphicx}
\usepackage{amsmath}
\usepackage{booktabs}
\usepackage{algorithm}
\usepackage{algorithmic}
\title{Secure Federated Matrix Factorization}
\author{
	Di Chai$^{1,2,*}$\and
	Leye Wang$^{3,*}$\and
	Kai Chen$^1$\and
	Qiang Yang$^{1,4}$
	\affiliations
	$^1$Hong Kong University of Science and Technology, China;
	$^2$Clustar, China\\
	$^3$Peking University, China;
	$^4$WeBank, China\\
	\emails
	dchai@connect.ust.hk, leyewang@pku.edu.cn, kaichen@cse.ust.hk, qyang@cse.ust.hk\\
	\small $^*$\textit{equal contribution, ranked alphabetically}
}
\begin{document}
	
	\maketitle
	
	\begin{abstract}
		To protect user privacy and meet law regulations, federated (machine) learning is obtaining vast interests in recent years. The key principle of federated learning is training a machine learning model without needing to know each user's personal raw private data. In this paper, we propose a secure matrix factorization framework under the federated learning setting, called \textit{FedMF}. First, we design a user-level distributed matrix factorization framework where the model can be learned when each user only uploads the gradient information (instead of the raw preference data) to the server. While gradient information seems secure, we prove that it could still leak users' raw data. To this end, we enhance the distributed matrix factorization framework with homomorphic encryption. We implement the prototype of FedMF and test it with a real movie rating dataset. Results verify the feasibility of FedMF. We also discuss the challenges for applying FedMF in practice for future research.
	\end{abstract}

\section{Introduction}

With the prevalence of government regulations and laws on privacy protection in the big data era (e.g., General Data Protection Regulation\footnote{\scriptsize https://eur-lex.europa.eu/legal-content/EN/TXT/?uri=CELEX:32016R0679R(02)}), privacy-preserving machine learning has obtained rapidly growing interests in both academia and industry. Among various techniques to achieve privacy-preserving machine learning, federated (machine) learning (FL) recently receives high attention. The original idea of FL was proposed by Google \cite{konevcny2016federated}, which targets at learning a centered model based on the personal information distributed at each user's mobile phone. More importantly, during model training, no user's raw personal information is transferred to the central server, thus ensuring the privacy protection. Also, the learned privacy-preserving model can be proved to hold almost similar predictive power compared to the traditional model learned on users' raw data. This highlights the practicality of FL as little predictive accuracy is sacrificed, especially compared to other accuracy-lossy privacy preserving mechanisms such as differential privacy \cite{dwork2011differential}.

Starting from the original Google paper, many researchers have been devoted into this promising and critical area. Recently, a nice survey paper on FL has been published \cite{yang2019federated}. While many research works have been done on FL, a popular machine learning technique, \textit{matrix factorization} (MF) \cite{koren2009matrix}, is still under-investigated in FL. Since MF is one of the prominent techniques widely employed in various applications such as item recommendation \cite{koren2009matrix} and environment monitoring \cite{wang2016sparse}, we highly believe that studying MF under FL is urgently required. This work is one of the pioneering research efforts toward this direction.


Taking recommendation systems as an example, two types of users' private information are leaked in traditional MF \cite{koren2009matrix}: (i) \textit{users' raw preference data}, and (ii) \textit{users' learned latent feature vectors}. As revealed by previous studies, either raw preference data or latent features can leak users' sensitive attributes, e.g., age, relationship status, political views, and sex orientations \cite{yang2016privcheck,kosinski2013private}. This highlights the importance of protecting users' private information during MF. Prior studies have studied privacy-preserving MF in two main types:

(1) \textbf{Obfuscation-based methods} obfuscate users' preference raw data before releasing it to the central server so as to ensure certain level of privacy protection (e.g., differential privacy) \cite{berlioz2015applying}. The pitfall is that obfuscation inevitably leads to the loss of predictive power of the learned latent feature vectors. Hence, these methods usually need to make a trade-off between the privacy protection and the model performance.

(2) \textbf{Encryption-based methods} use advanced encryption schemes such as homomorphic encryption for implementing privacy-preserving MF \cite{kim2016efficient}. While they usually do not need to sacrifice predictive power for privacy protection, they commonly require a third-party crypto-service provider. This makes the system implementation complicated as such a provider is not easy to find in practice.\footnote{Some studies suggest that governments can perform this role, but this is still not the case in reality now \cite{kim2016efficient}.} Moreover, if the crypto-service provider collude with the recommendation server, then no user privacy protection can be preserved \cite{kim2016efficient,nikolaenko2013privacy}.

This research proposes a novel FL-based MF framework, called \textbf{FedMF} (Federated Matrix Factorization). FedMF employs distributed machine learning and homomorphic encryption schemes. In brief, FedMF lets each user compute the gradients of his/her own rating information locally and then upload the gradients (instead of the raw data) to the server for training. To further enhance security, each user can encrypt the gradients with homomorphic encryption. With FedMF, two shortcomings of the traditional obfuscation- or encryption-based methods can be addressed: (i) no predictive accuracy is lost as we do not obfuscate data; (ii) no third-party crypto-service provider is required as each user's device can handle the secure gradient computing task.

In summary, we have the following contributions:

(1) To the best of our knowledge, we design the first FL-based secure MF framework, called \textit{FedMF}, which can overcome the shortcomings of traditional obfuscation- or encryption-based mechanisms as aforementioned.

(2) To implement FedMF, first, we design a user-level distributed matrix factorization framework where the model can be learned when each user only uploads the gradient information (instead of the raw preference data) to the server. Though gradient information seems secure, we prove that it could still leak users' raw data to some extent. Then, we enhance the distributed matrix factorization framework with homomorphic encryption to increase the security.

(3) We implement a prototype of FedMF (the code will be published in https://github.com/Di-Chai/FedMF). We test the prototype on a real movie rating dataset. Results verify the feasibility of FedMF.

It is worth noting that, similar to FedMF, \cite{DBLP:journals/corr/abs-1901-09888} tried to develop a federated collaborative filtering system. However, \cite{DBLP:journals/corr/abs-1901-09888} directly let users upload their gradient information to the server. As we will prove in this paper, gradients can still reveal users' original preference data. Hence, a more secure system like FedMF is required to rigorously protect users' privacy.

\section{Preliminaries}

In this section, we briefly introduce two techniques closely related to our research, \textit{horizontal federated learning} and \textit{additively homomorphic encryption}.

\subsection{Horizontal Federated Learning} \label{HFL}

\textit{Federated learning} is a method that enables a group of data owners to train a machine learning model on their joint data and nobody can learn the data of the participants. Federated learning can be categorized based on the distribution characteristics of the data ~\cite{yang2019federated}. One category of federated learning is \textit{horizontal federated learning}. Horizontal federated learning is introduced in the scenarios when the data from different contributors share the same feature space but vary in samples. In our secure matrix factorization recommendation system, the rating information distributed on each user's device has exactly the same feature space, while different users are exactly different samples. So our matrix factorization in federated learning can be categorized as horizontal federated learning. Following the typical assumption of horizontal federated learning~\cite{yang2019federated}, we assume \textbf{all the users are honest and the system is designed to be secure against an honest-but-curious server}.

\subsection{Additively Homomorphic Encryption} \label{HE}

\textit{Homomorphic encryption} (HE) is often used in federated learning to protect user's privacy by providing encrypted parameter exchange. HE is a special kind of encryption scheme that allows any third party to operate on the encrypted data without decrypting it in advance. An encryption scheme is called homomorphic over an operation '$\star$' if the following equation holds:
\begin{equation}
E(m_1) \star E(m_2) = E(m_1 \star m_2), \ \forall m_1,m_2 \in M
\end{equation}
where $E$ is an encryption algorithm and $M$ is the set of all possible messages \cite{acar2018survey}.

\textit{Addictively homomorphic encryption} is homomorphic over addition. Typically, it consists of the following functions:

\begin{itemize}
	\item $KeyGen \rightarrow (pk, sk)$: key generation process, where $pk$ is the public key and $sk$ is the secret key.
	\item $Enc(m, pk) \rightarrow c$: encryption process, where $m$ is the message to be encrypted and c is the ciphertext.
	\item $Dec(c, sk) \rightarrow m$: decryption process
	\item $Add(c_1, c_2) \rightarrow c_{a}(i.e. \ Enc(c_1 + c_2))$: add operation on ciphertext, $c_a$ is the ciphertext of plaintexts' addition.
	\item $DecAdd(c_a, sk) \rightarrow m_a$: decrypt $c_a$, getting the addition of plaintexts. 
\end{itemize}

\section{User-level Distributed Matrix Factorization}
\label{sec:distbuted_mf}

We firstly introduce the matrix factorization optimization method used in our paper, stochastic gradient descent \cite{koren2009matrix}. Based on it, we design a user-level distributed matrix factorization framework.

\subsection{Stochastic Gradient Descent}

Suppose we have $n$ users, $m$ items and each user rated a subset of $m$ items. For $[n] := \{1,2...,n\}$ as the set of users and $[m]:=\{1,2,...,m\}$ as the set of items, we denote $\mathcal{M} \in [n] \times [m] $ as user-item rating pairs which a rating has been generated, $M=|\mathcal{M}|$ as the total number of ratings and $r_{i,j}$ represents the rating generated by user $i$ for item $j$.

Given the rating information ${r_{ij}:(i,j) \in \mathcal{M}}$, the recommendation systems are expected to predict the rating values of all the items for all the users. Matrix factorization formulates this problem as fitting a bi-linear model on the existing ratings. In particular, user profile matrix $U \in \mathbf{R} ^ {n \times d}$ and item profile matrix $V \in \mathbf{R} ^ {m \times d}$ are computed, the resulting profile matrices are used to predict user $i$'s rating on item $j$, which is $ \langle u_i, v_j \rangle $. The computing process of $U$ and $V$ can be done by solving the following regularized least squares minimization:
\begin{equation}
\min_{U,V}\frac{1}{M}(r_{i,j}- \langle u_i,v_j \rangle )^2+\lambda||U||_2^2+\mu||V||_2^2
\end{equation}
where $\lambda$ and $\mu$ are small positive values to rescale the penalizer. Stochastic gradient descent iteratively updates $U$ and $V$ with the following equations \cite{koren2009matrix}:
\begin{equation}u_i^t=u_i^{t-1}-\gamma\bigtriangledown_{u_i}F(U^{t-1},V^{t-1})
\label{eq:user_update}
\end{equation}
\vspace{-1em}
\begin{equation}v_i^t=v_i^{t-1}-\gamma\bigtriangledown_{v_i}F(U^{t-1},V^{t-1})
\label{eq:server_update}
\end{equation}
where
\begin{equation}\bigtriangledown_{u_i}F(U,V)=-2\sum_{j:(i,j)}v_j(r_{ij}- \langle u_i,v_j \rangle )+2\lambda u_i\end{equation}
\vspace{-1em}
\begin{equation}\bigtriangledown_{v_j}F(U,V)=-2\sum_{i:(i,j)}u_i(r_{ij}- \langle u_i,v_j \rangle )+2\lambda v_j\end{equation}
The number of iterations relies on the stopping criteria. A typical criteria is to set a small threshold $\varepsilon$, such that the training stops when the gradient $\bigtriangledown_{u_i}F$ and $\bigtriangledown_{v_j}F$ (or one of them) are smaller than $\varepsilon$. 

\subsection{Distributed Matrix Factorization}

In the distributed matrix factorization scenario, users hold their rating information locally and the model is trained on their joint data. To achieve this goal, we leverage a distributed matrix factorization method, which decomposes the iterative updating process into two parts that are performed on the user side and the server side, respectively. In particular, equation~\eqref{eq:user_update} is executed on user $i$'s device, namely \textit{user update}, and equation~\eqref{eq:server_update} is performed on the server, called \textit{server update}. This decomposition prevents the server from directly knowing users' raw preference data or learned profiles.

Algorithm \ref{alg:algorithm1} shows our user-level distributed matrix factorization method. The server keeps providing the latest item profile matrix $V$ for all the users to download. Having the latest downloaded $V$ and his/her own rating information, each user $i$ performs local updates and computes $Gradient_i$, which will be sent back to the server to update item profiles.

\begin{algorithm}[tb]
	\caption{User-level Distributed Matrix Factorization}
	\label{alg:algorithm1}
	\textbf{Init}: Server initializes item profile matrix V \\
	\textbf{Init}: User initializes user profile matrix U \\        
	\textbf{Output}: Converged U and V
	\begin{algorithmic}
		\STATE \textbf{Server keeps latest item-profile for all users' download}
		\STATE \textbf{User local update:}
		\STATE \hspace{\algorithmicindent} Download $V$ from server, peform local updates:
		\STATE \hspace{\algorithmicindent} $u_i^t=u_i^{t-1}-\gamma\bigtriangledown_{u_i}F(U^{t-1},V^{t-1})$
		\STATE \hspace{\algorithmicindent} $Gradient_i = \gamma\bigtriangledown_{v_i}F(U^{t-1},V^{t-1}) $
		\STATE \hspace{\algorithmicindent} Send $Gradient_i$ to server
		\STATE \textbf{Server update:}
		\STATE \hspace{\algorithmicindent} Receive $Gradient_i$ from user-i
		\STATE \hspace{\algorithmicindent} Perform update : $v_i^t=v_i^{t-1}- Gradient_i$
	\end{algorithmic}
\end{algorithm}

\section{Gradients Leak Information} \label{section-GLI}

Algorithm \ref{alg:algorithm1} shows a framework that allows the server to build a matrix factorization recommendation system on a distributed dataset, i.e. users keep rating information locally. In this framework, users iteratively send $Gradient$ information to server in plain text. Next we are going to prove that such a distributed matrix factorization system cannot protect users' rating information against the server, i.e. server can deduce users' rating data using the $Gradient$.

For user-vector $u_i$, suppose the user rated item-set is $M_i$. We will have the following equation at time $t$
\begin{equation}
\label{s4-eq1}
u_i^t(r_{ij}- \langle u_i^t,v_j^t \rangle )=G_j^t, \ j \in M_i
\end{equation}
where $G$ is the gradient to be uploaded from the user $i$ to the server for updating item-profiles. Similarly, at time $t+1$,
\begin{equation}u_i^{t+1}(r_{ij}- \langle u_i^{t+1},v_j^{t+1} \rangle )=G_j^{t+1}, \ j \in M_i\end{equation}

The correlation between $U_t$ and $U_{t+1}$ is :

\begin{equation}
\label{s4-eq2}
u_i^t-u_i^{t+1}=-2\sum_{j\in M_i}v_j^t (r_{ij}- \langle u_i^t,v_j^t \rangle )
\end{equation}

Take a close look at the element-wise calculation of equation \eqref{s4-eq1}--\eqref{s4-eq2}, where we denote the latent user and item vectors ($u_i$ and $v_j$) are $D$ dimension:
\begin{equation}
\label{s4-eq3}
\begin{cases}
u^t_{i1}(r_{ij}-\sum_{m=1}^{D}u^t_{im}v^t_{jm})=G_{j1}^t \\
\vdots \\
u^t_{ik}(r_{ij}-\sum_{m=1}^{D}u^t_{im}v^t_{jm})=G_{jk}^t \\
\vdots \\
u^t_{iD}(r_{ij}-\sum_{m=1}^{D}u^t_{im}v^t_{jm})=G_{jD}^t \\
\end{cases}
\end{equation}

\begin{equation}
\label{s4-eq4}
\begin{cases}
u^{t+1}_{i1}(r_{ij}-\sum_{m=1}^{D}u^{t+1}_{im}v^{t+1}_{jm})=G_{j1}^{t+1} \\
\vdots \\
u^{t+1}_{ik}(r_{ij}-\sum_{m=1}^{D}u^{t+1}_{im}v^{t+1}_{jm})=G_{jk}^{t+1} \\
\vdots \\
u^{t+1}_{iD}(r_{ij}-\sum_{m=1}^{D}u^{t+1}_{im}v^{t+1}_{jm})=G_{jD}^{t+1} \\
\end{cases}
\end{equation}

\begin{equation}
\label{s4-eq5}
\begin{cases}
u^t_{i1} - u^{t+1}_{i1} = -2\sum^{N}_{n=1} v^t_{n1}(r_{in}-\sum_{m=1}^{D}u^t_{im}v^t_{nm}) \\
\vdots \\
u^t_{ik} - u^{t+1}_{ik} = -2\sum^{N}_{n=1} v^t_{nk}(r_{in}-\sum_{m=1}^{D}u^t_{im}v^t_{nm}) \\
\vdots \\
u^t_{iD} - u^{t+1}_{iD} = -2\sum^{N}_{n=1} v^t_{nD}(r_{in}-\sum_{m=1}^{D}u^t_{im}v^t_{nm}) \\
\end{cases}
\end{equation}

Now we turn to analyze the $k$-th entry of $u_i$, $u_{ik}$.
From equation \eqref{s4-eq3}, we have :

\begin{equation}
\label{s4-eq6}
\frac{u^t_{ik}}{u^t_{i(k+1)}} = \frac{G^t_{ik}}{G^t_{i(k+1)}}
\end{equation}
\begin{equation}
\label{s4-eq7}
r_{ij}-\sum_{m=1}^{D}u^t_{im}v^t_{jm}=\frac{G_{jk}^t}{u^t_{ik}}
\end{equation}

Plug equation \eqref{s4-eq6} into \eqref{s4-eq4}, we will have
\begin{equation}
\label{s4-eq8}
u^t_{ik} - u^{t+1}_{ik} =  - 2 \frac{1}{u^t_{ik}} \sum^{N}_{n=1} v^t_{nk}G_{nk}^t
\end{equation}

Thus we can represent $u^{t+1}_{ik}$ using $u^{t}_{ik}$ as:
\begin{equation}
\label{s4-eq9}
u^{t+1}_{ik} = u^t_{ik} + 2 \frac{1}{u^t_{ik}} \sum^{N}_{n=1} v^t_{nk}G_{nk}^t
\end{equation}

From equation \eqref{s4-eq3} and \eqref{s4-eq4} we have:
\begin{equation}
\label{s4-eq10}
\frac{G_{jk}^t}{u^t_{ik}}+\sum_{m=1}^{D}u^t_{im}v^t_{jm}=\frac{G_{jk}^{t+1}}{u^{t+1}_{ik}}+\sum_{m=1}^{D}u^{t+1}_{im}v^{t+1}_{jm}
\end{equation}

Plug equation \eqref{s4-eq9} into \eqref{s4-eq10}:
\begin{align}
\label{s4-eq11}
& \frac{G_{jk}^t}{u^t_{ik}} + \sum_{m=1}^{D}u^t_{im}v^t_{jm} = \frac{G_{jk}^{t+1}}{u^t_{ik} + 2 \frac{1}{u^t_{ik}} \sum^{N}_{n=1} v^t_{nk}G_{nk}^t} + \nonumber\\
& \sum_{m=1}^{D}(u^t_{im} + 2 \frac{1}{u^t_{im}} \sum^{N}_{n=1} v^t_{nm}G_{nm}^t)v^{t+1}_{jm}	
\end{align}
which is,
\begin{align}
\label{s4-eq11}
& \frac{G_{jk}^t}{u^t_{ik}} - \frac{G_{jk}^{t+1}}{u^t_{ik} + 2 \frac{1}{u^t_{ik}} \sum^{N}_{n=1} v^t_{nk}G_{nk}^t} \nonumber\\
& =\sum_{m=1}^{D}[(u^t_{im} + 2 \frac{1}{u^t_{im}} \sum^{N}_{n=1} (v^t_{nm}G_{nm}^t))v^{t+1}_{jm}-u^t_{im}v^t_{jm}]
\end{align}
Let $\alpha_k = 2\sum^{N-1}_{n=0} v^t_{nk}G_{nk}^t$,
\begin{align}
\label{s4-eq12}
\frac{G_{jk}^t}{u^t_{ik}} - \frac{G_{jk}^{t+1}}{u^t_{ik} + \frac{\alpha_k}{u^t_{ik}}} & = \sum_{m=1}^{D}[(u^t_{im} + \frac{\alpha_m}{u^t_{im}})v^{t+1}_{jm}-u^t_{im}v^t_{jm}] \nonumber\\
& = \sum_{m=1}^{D}[(v^{t+1}_{jm}-v^{t}_{jm})u^t_{im}+\frac{\alpha_m v^{t+1}_{jm}}{u^t_{im}}]
\end{align}
From equation \eqref{s4-eq6}, we can have:
\begin{equation}
\label{s4-eq13}
u^t_{im} = \frac{G^t_{jm}}{G^t_{jk}} u^t_{ik}
\end{equation}
Plug equation \eqref{s4-eq13} into \eqref{s4-eq12}:
\begin{align}
\label{s4-eq14}
& \frac{G_{jk}^t}{u^t_{ik}} - \frac{G_{jk}^{t+1}}{u^t_{ik} + \frac{\alpha_k}{u^t_{ik}}}  \nonumber\\
& = \sum_{m=1}^{D}[(v^{t+1}_{jm}-v^{t}_{jm})\frac{G^t_{jm}}{G^t_{jk}} u^t_{ik}+\frac{\alpha_m v^{t+1}_{jm}}{\frac{G^t_{jm}}{G^t_{jk}} u^t_{ik}}]  \nonumber\\
& = \frac{u^t_{ik}}{G^t_{jk}} \sum_{m=1}^{D}[(v^{t+1}_{jm}-v^{t}_{jm})G^t_{jm}] + \frac{G^t_{jk}}{u^t_{ik}} \sum_{m=1}^{D}[\frac{\alpha_m v^{t+1}_{jm}}{G^t_{jm}}]
\end{align}
Denote $\beta_j$ and $\gamma_j$ as follow:
\begin{equation}
\begin{cases}
\beta_j=\sum_{m=1}^{D}[(v^{t+1}_{jm}-v^{t}_{jm})G^t_{jm}] \\
\gamma_j=\sum_{m=1}^{D}[\frac{\alpha_m v^{t+1}_{jm}}{G^t_{jm}}]
\end{cases}
\end{equation}
We will have:
\begin{equation}
\label{s4-eq15}
\frac{G_{jk}^t}{u^t_{ik}} - \frac{G_{jk}^{t+1}}{u^t_{ik} + \frac{\alpha_k}{u^t_{ik}}} = \frac{u^t_{ik}}{G^t_{jk}} \beta_j + \frac{G^t_{jk}}{u^t_{ik}} \gamma_j
\end{equation}

Since we know there must be one real scalar of $u^t_{ik}$ that satisfies equation \eqref{s4-eq15}. We can use some iterative methods to compute a numeric solution of \eqref{s4-eq15}, e.g., Newton's method.

After getting $u^t_i$, we can use equation \eqref{s4-eq3} to compute $r_i$, which can be written as:
\begin{equation}
r_{ij}=\frac{G_{jk}^t}{u^t_{ik}}+\sum_{m=1}^{D}u^t_{im}v^t_{jm}
\end{equation}

In summary, knowing the gradients of a user uploaded in two continuous steps, we can infer this user's rating information. Thus, we propose a secure matrix factorization framework based on homomorphic encryption, which will be elaborated in the next section.

\begin{figure*}[t]
	\centering
	\includegraphics[width=.7\linewidth]{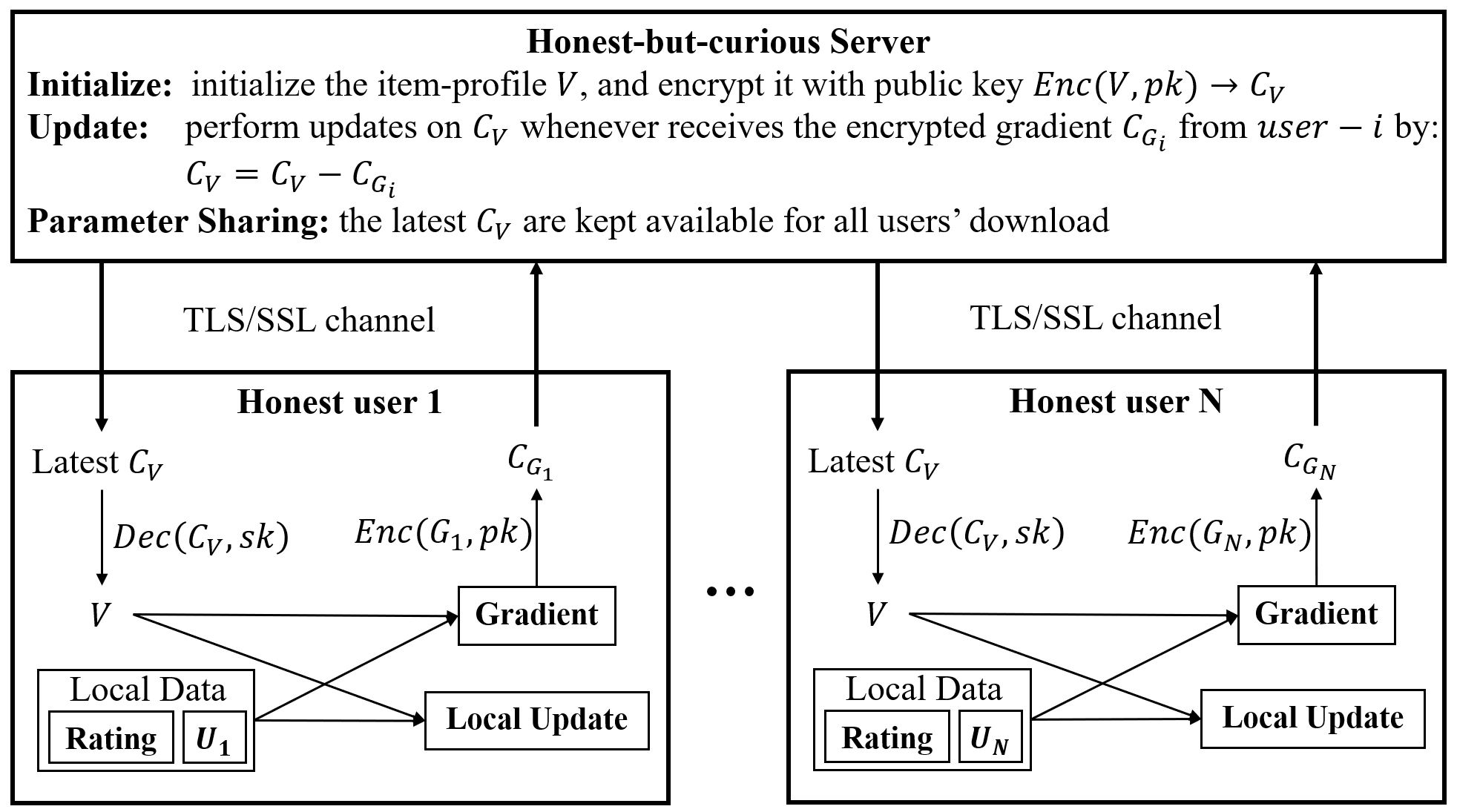}
	\caption{Overview of FedMF}
	\label{fig:framework}
\end{figure*}

\section{FedMF: Federated Matrix Factorization}

To overcome this information leakage problem, we propose to encode the gradients such that server cannot inverse the encoding process. Then, the encoded data leaks no information. Meanwhile, the server should still be able to perform updates using the encoded gradients. One way to achieve such a goal is using homomorphic encryption.

Figure \ref{fig:framework} shows a framework of our method, called \textit{FedMF} (Federated Matrix Factorization). Two types of participants are involved in this framework, the server and the users. As previously illustrated in Sec. \ref{HFL}, we assume that the server is honest-but-curious, the users are honest, and the privacy of the users is protected against the server. 

\textbf{Key Generation:} As the typical functions involved in homomorphic encryption (Sec. \ref{HE}), we first generate the \textit{public key} and \textit{secret key}. The key generation process is carried out on one of the users. The \textit{public key} is known to all the participants including the server. And the \textit{secret key} is only shared between users and needs to be protected against the server. After the keys are generated, different TLS/SSL secure channels will be established for sending the \textit{public key} and \textit{secret key} to the corresponding participants.

\textbf{Parameter Initialization:} Before starting the matrix factorization process, some parameters need to be initialized. The item profile matrix is initialized at the server side while the user profile matrix is initialized by each user locally. 

\textbf{Matrix Factorization:} Major steps include,
\begin{enumerate}
	\item The server encrypts item profile $V$ using \textit{public key}, getting the ciphertext $C_V$. From now on, the latest $C_V$ is prepared for all users' download.
	\item Each user downloads the latest $C_V$ from the server, and decrypts it using \textit{secret key}, getting the plaintext of $V$. $V$ is used to perform local update and compute the gradient $G$. Then $G$ is encrypted using \textit{public key}, getting ciphertext $C_V$. Then a TLS/SSL secure channel is built, $C_V$ is sent back to the server via this secure channel.
	\item After receiving a user's encrypted gradient, the server updates the item profile using this ciphertext : $C_V^{t+1} = C_V^t - C_G$. Afterwards, the latest $C_V$ is prepared for users' downloading.
	\item Step 2 and 3 are iteratively executed until convergence.
\end{enumerate}

\textit{Security against Server:} As shown in Fig.~\ref{fig:framework}, only ciphertext is sent to the server in FedMF. So no bit of information will be leaked to the  server as long as our homomorphic encryption system ensures ciphertext indistinguishability against chosen plaintext attacks \cite{goldreich2009foundations}.

\textit{No Accuracy Decline:} We also claim that FedMF is accuracy equivalent to the user-level distributed matrix factorization. This is because the parameter updating process is same as the distributed matrix factorization (Sec.~\ref{sec:distbuted_mf}) if the homomorphic encryption part is removed.

\section{Prototype and Evaluation}

In this section, we choose Paillier encryption method \cite{paillier1999public} to instantiate a prototype of our system and use a real movie rating dataset to evaluate the prototype.

\subsection{Prototype Implementation}

We use Paillier encryption \cite{paillier1999public} to build a prototype of FedMF. Paillier encryption is a probabilistic encryption schema based on composite residuosity problem \cite{jager2012generic}. Given \textit{publick key} and encrypted plaintext, paillier encryption has the following homomorphic property operations:

	\textit{op1.} $E(m_1) \cdot E(m_2) \ (mode \ n^2) = E(m_1+m_2 \ (mode \ n))$
	
	\textit{op2.} $E(m_1) \cdot g^{m_2} \ (mode \ n^2) = E(m_1+m_2 \ (mode \ n))$
	
	\textit{op3.} $E(m_1)^{m_2} \ (mode \ n^2) = E(m_1m_2 \ (mode \ n))$

Typically, paillier encryption requires the plaintext to be positive integer. But in our system, the data are all in the form of floating point numbers and some of them might be negative. Thus we need to extend the encryption method to support our system. 

\textbf{Float}. In brief, a base exponent was multiplied to the decimal, the integer part of the multiplication result $I$ and the base exponent $e$ was treated as the integer representation of the floating point number, i.e. $(I, e)$. In the encryption process, only $I$ is encrypted, the ciphertext will be $(C_I, e)$. Then the ciphertext with the same $e$ can directly conduct operation \textit{op1} to get the encrypted summation of plaintext, ciphertext with different $e$ needed to recalibrate such that $e$ is the same. In practice, we use the same base exponent such that no information will leak from $e$. 

\textbf{Negative number.} A \textit{max number} parameter is set to handle the negative numbers. The \textit{max number} can be set to half of $n$ in \textit{pk}, which means we assume all of our data is smaller than \textit{max number}, such an assumption is easy to satisfy since $n$ is usually set to a very large prime number. Then we perform mode $n$ on all the plaintext, all the positive number have no changes and all the negative numbers become positive numbers greater than  \textit{max number}. In the decryption process, if the decrypted plaintext is greater than \textit{max number}, we minus $n$ to get the correct negative plaintext.

\textbf{FullText or PartText.} Usually the rating or feedback comprises a sparse matrix \cite{koren2009matrix} which means the amount of feedback from a user could be very limited. Therefore, two different settings are implemented our system. Both of them follow the overall steps of FedMF, but are slightly different at the user uploading process. In one setting called \textit{FullText}, users upload gradients for all the items; the gradient is set to 0 if a user does not rate an item. In the other setting called \textit{PartText}, users only upload the gradients of the rated items. They both have advantages and disadvantages, \textit{PartText} leaks information about which items the user has rated but has higher computation efficiency, \textit{FullText} leaks no information but needs more computation time.

We utilize an open source python package, \textit{python-paillier}\footnote{\url{https://github.com/n1analytics/python-paillier}} to accomplish the encryption part in our prototype system. 

\subsection{Evaluation}

\textbf{Dataset:} To test the feasibility of our system, we use a real movie rating dataset \cite{harper2016movielens} from MovieLens which contains 100K rating information made by 610 users on 9724 movies. This dataset is also used in other homomorphic-encrypted MF works such as \cite{nikolaenko2013privacy} and \cite{kim2016efficient}.

\textbf{Parameters:} In Paillier encryption, we set the length of \textit{public key} to 1024. The bandwidth of communication is set to \textit{1 Gb/s}. In the matrix factorization process, we set the dimension of user and item profile to 100. 

\textbf{Environment:} All the test experiments are performed on a server with 5.0GHz 6-core CPU and 32GB RAM, where the operation system is Windows and the program language is Python. We used a module called \textit{gmpy} \footnote{\textit{gmpy} is a c-coded Python extension module that supports multiple-precision arithmetic, \url{https://github.com/aleaxit/gmpy}} to accelerate the homomorphic encryption part in Python such that it is as fast as C++ implementation.

\begin{table}[t]
	\centering
	\footnotesize
	\begin{tabular}{llll}
		\hline
		\textbf{\#Item}   & \textbf{\#Rating}  & \textbf{PartText} & \textbf{FullText} \\
		\hline
		40       & 8307   & 34.39  & 90.94  \\
		50       & 9807   & 44.05  & 113.34  \\
		60       & 11214  & 46.34  & 141.52  \\
		80       & 13817  & 52.91  & 182.27  \\
		160      & 22282  & 92.81  & 374.85  \\
		320      & 34172  & 140.51 & 725.72  \\
		640      & 49706  & 178.24 & 1479.40  \\
		1280     & 67558  & 264.10 & 2919.91  \\
		2560     & 83616  & 334.79 & 5786.01  \\
		\hline
	\end{tabular}
	\caption{Time consumption of each iteration (seconds).}
	\label{tab:result1}
\end{table}

\begin{figure}[t]
	\includegraphics[width=.9\linewidth]{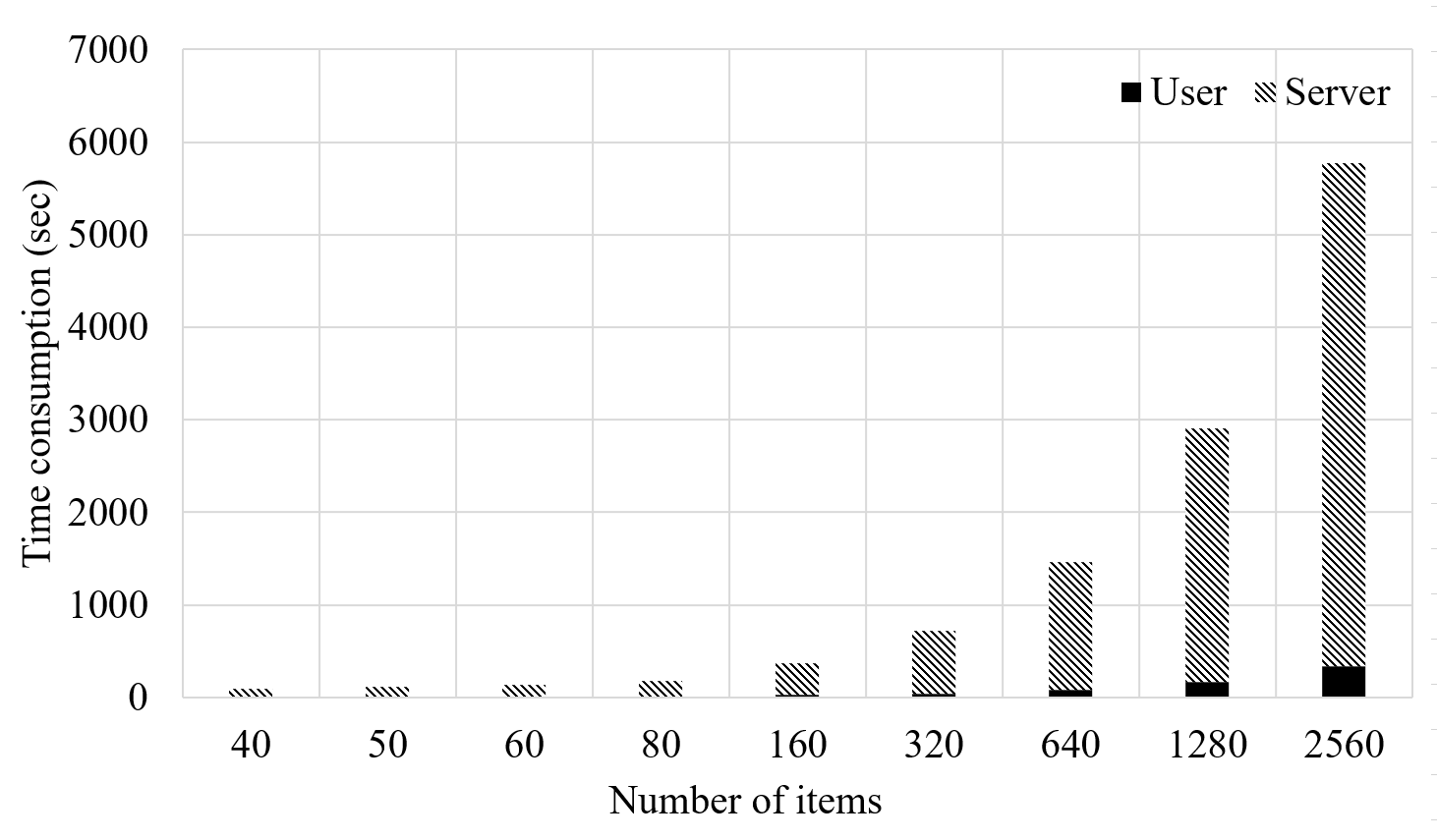}
	\vspace{-.7em}
	\caption{User-Server time consumption ratio of FullText}
	\label{fig:User-Server-FT}
\end{figure}

\begin{figure}[t]
	\includegraphics[width=.9\linewidth]{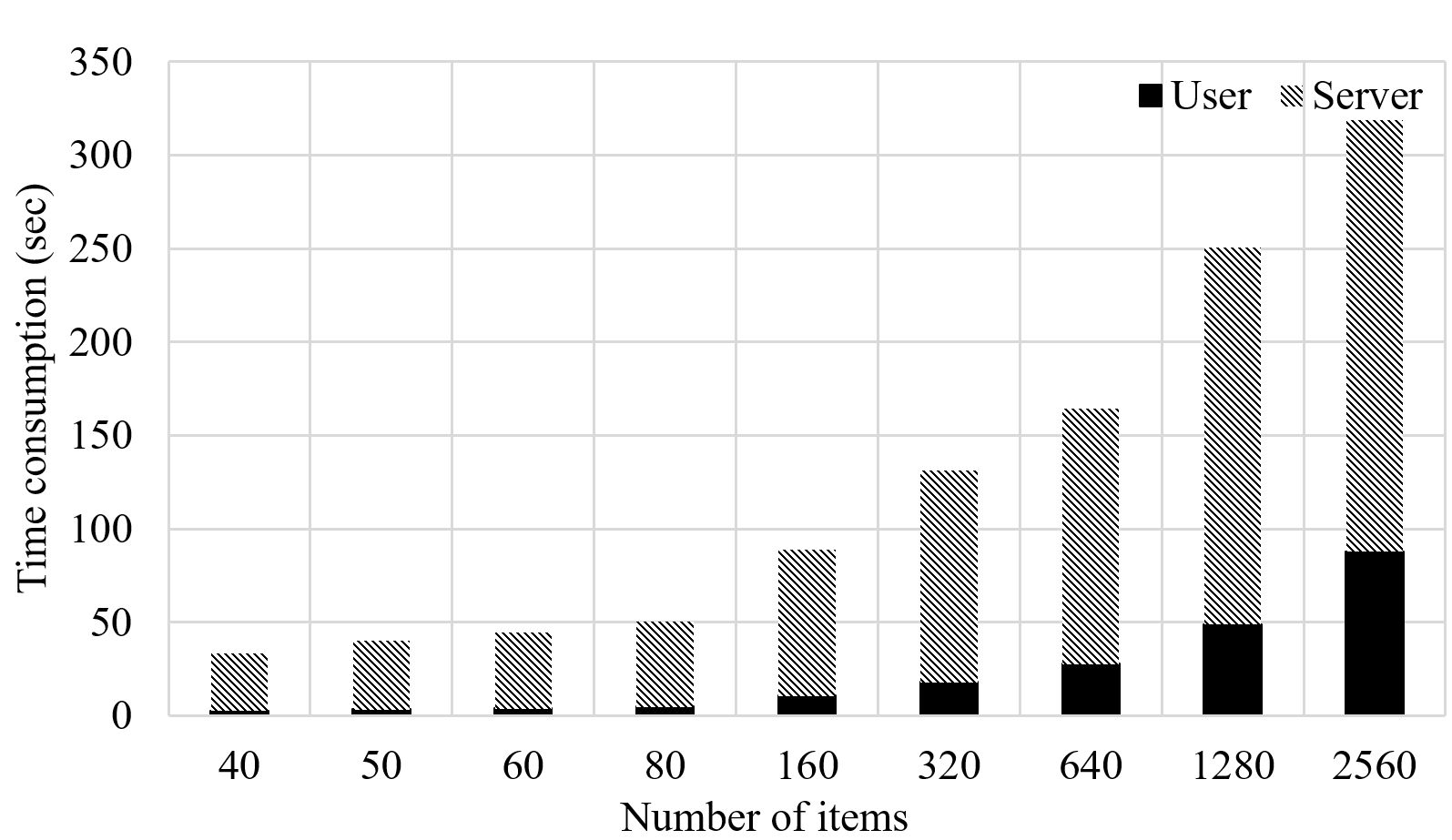}
	\vspace{-.7em}
	\caption{User-Server time consumption ratio of PartText}
	\label{fig:User-Server-PT}
\end{figure}

\textbf{Performance}: Since neither distributed computing or homomorphic encryption mechanisms will affect the computation values, FedMF will output the same user and item profiles as the original MF algorithm. Hence, the major objective of the experiments is testing the computation time of FedMF.
Fixing the number of users to 610, Table \ref{tab:result1} shows the time consumption of each iteration of \textit{PartText} and \textit{FullText} (one iteration means  all of the 610 users' uploaded gradients are used to update the item profiles once). For both \textit{PartText} and \textit{FullText}, the time consumption is quite good when there are not too many items, and the time efficiency decreases when more items are given. Roughly, the time consumed for each iteration linearly increases with the number of items. Compared with \textit{Fulltext}, \textit{PartText} is more efficient but it leaks some information. Particularly, \textit{PartText} is nearly 20 times faster than the \textit{Fulltext} solution.

Fig.~\ref{fig:User-Server-FT} and \ref{fig:User-Server-PT} show the ratio of the user and server updating time when the number of items changes. The communication time is dismissed from the figures because it is too small compared with the user and server updating time. For example,  $\sim$80MB of gradient data need to be sent to server when the item number is 2560 and it will cost only 1.25 seconds. From these figures, we can find out that  $\sim$95\% of time in one iteration is spent on server updates, which means if we increase the computing power of the server or improve the homomorphic encryption method such that the complexity of computation on ciphertext is lowered, the time efficiency of the whole system will improve significantly. This would be our future work.

\section{Conclusion and Future Work}

In this paper, we propose a novel secure matrix factorization framework in federated machine learning, called \textit{FedMF}. More specifically, we first prove that a distributed matrix factorization system where users send gradients to the server in forms of plaintext will leak users' rating information. Then, we design a homomorphic encryption based secure matrix factorization framework. We have proved that our system is secure against an honest-but-curious server, and the accuracy is same as the matrix factorization on users' raw data.

Experiments on real-world data show that FedMF's time efficiency is acceptable when the number of items is small. Also note that our system's time consumption linearly increases with the number of items. To make FedMF more practical in reality, we still face several challenges:

\textit{More efficient homomorphic encryption.} As we have discussed before, about 95\% of our system's time consumption is spent on server updates, where the computation is all performed on the ciphertext. If we can improve the homomorphic encryption's efficiency when conducting operations on ciphertext, our system's performance will increase.

\textit{Between \textit{FullText} and \textit{PartText}.} Our experiments have shown that \textit{PartText} is much more efficient than \textit{FullText}, but \textit{PartText} reveals the set of items rated by a user. This information, without the exact rating scores, may still leak users' sensitive information \cite{yang2016privcheck}. Perhaps we can ask users to upload more gradients than only the rated items, but not all the items, so as to increase efficiency compared to \textit{FullText}, while not leaking the exactly rated item set.

\textit{More secure definitions.} Currently, we use a typical horizontal federated learning secure definition, which assumes honest participants and an honest-but-curious server. Next, we can explore more challenging secure definitions, such as how to build a secure system where the server is honest-but-curious, and some participants are malicious and the malicious participants may collude with the server.

\bibliographystyle{named}
\bibliography{ref.bib}

\end{document}